%% file: main.tex
\documentclass{article}

\usepackage{arxiv}

\usepackage[utf8]{inputenc} 
\usepackage[T1]{fontenc}    
\usepackage{hyperref}       
\usepackage{url}            
\usepackage{booktabs}       
\usepackage{amsfonts}       
\usepackage{nicefrac}       
\usepackage{microtype}      
\usepackage{lipsum}
\usepackage{graphicx}

\usepackage{cite}
\usepackage{amsmath,amssymb,amsfonts}
\usepackage{algorithmic}
\usepackage{graphicx}
\usepackage{textcomp}
\usepackage{bm}
\usepackage{array}
\usepackage{xspace}
\usepackage{xcolor}
\usepackage{booktabs}
\usepackage{ragged2e}
\usepackage{multirow}
\usepackage{mathtools}
\usepackage{footnote}
\usepackage{subcaption}
\usepackage{afterpage}
\usepackage{hyperref}
\usepackage{relsize}
\usepackage{color,soul}
\usepackage{paralist}
\usepackage{balance}

\newcommand{\eg}{e.g.,\xspace}

\newcommand{\etal}{\emph{et al.}\xspace}

\newcommand\tref[1]{\ref{table:#1}}
\newcommand\sref[1]{\ref{sec:#1}}
\newcommand\fref[1]{\ref{fig:#1}}

\newcommand\tlab[1]{\label{table:#1}}
\newcommand\slab[1]{\label{sec:#1}}
\newcommand\flab[1]{\label{fig:#1}}

\newcommand{\code}[1]{\texttt{#1}}

\def\BibTeX{{\rm B\kern-.05em{\sc i\kern-.025em b}\kern-.08em
    T\kern-.1667em\lower.7ex\hbox{E}\kern-.125emX}}

\begin{document}

\title{Automated Quality Assessment of Hand Washing Using Deep Learning}

\author{
 Maksims Ivanovs, Roberts Kadikis, Atis Elsts \\
  Institute of Electronics and Computer Science (EDI),\\
  Dzerbenes 14, LV-1006, Riga, Latvia
   \And
    Martins Lulla, Aleksejs Rutkovskis \\
  Riga Stradins University,\\
  Dzirciema iela 16, LV-1007, Riga, Latvia
  }





\maketitle

\begin{abstract}
Washing hands is one of the most important ways to prevent infectious diseases, including COVID-19.
Unfortunately, medical staff does not always follow the World Health Organization (WHO) hand washing guidelines in their everyday work.
To this end, we present neural networks for automatically recognizing the different washing movements defined by the WHO.
We train the neural network on a part of a large (2000+ videos) real-world labeled dataset with the different washing movements.
The preliminary results show that using pre-trained neural network models such as MobileNetV2 and Xception for the task, it is possible to achieve $>$64\,\% accuracy in recognizing the different washing movements.
We also describe the collection and the structure of the above open-access dataset created as part of this work.
Finally, we describe how the neural network can be used to construct a mobile phone application for automatic quality control and real-time feedback for medical professionals.
\end{abstract}


\input{01-introduction}
\input{02-relwork}
\input{03-datacollection}
\input{04-recognition}

\input{05-application}

\section{Conclusion}

The COVID-19 pandemic has reminded the world about the importance of simple everyday practices such as correctly washing hands. 
The work presented in this paper leads towards a system that is able to help medical professionals to perform this task better and more reliably.
To this end, we described neural network architectures for automatically recognizing the different washing movements defined by the WHO.
We use two pre-trained networks for the task: MobileNetV2 and Xception.
We train the networks on a large real-world dataset which we have collected and made available.
The networks show the following accuracy in the task of classifying the different movements: MobileNetV2 64\,\% and Xception 67\,\%.
In the future, a selected neural network will be integrated in a mobile application for hand washing quality control and feedback.

\section*{Acknowledgement}

This work was funded by the project VPP-COVID-2020/1-0004 ``Integration of reliable technologies for protection against Covid-19 in healthcare and high risk areas''.

We gratefully acknowledge RSU staff and students who contributed to the labor-intensive task of labeling of the handwashing video dataset.

~
~



\vfill


\end{document}

%% file: 01-introduction.tex
\section{Introduction}
\slab{intro}

\begin{figure}[h]
  \begin{center}
    \includegraphics[width=0.8\linewidth]{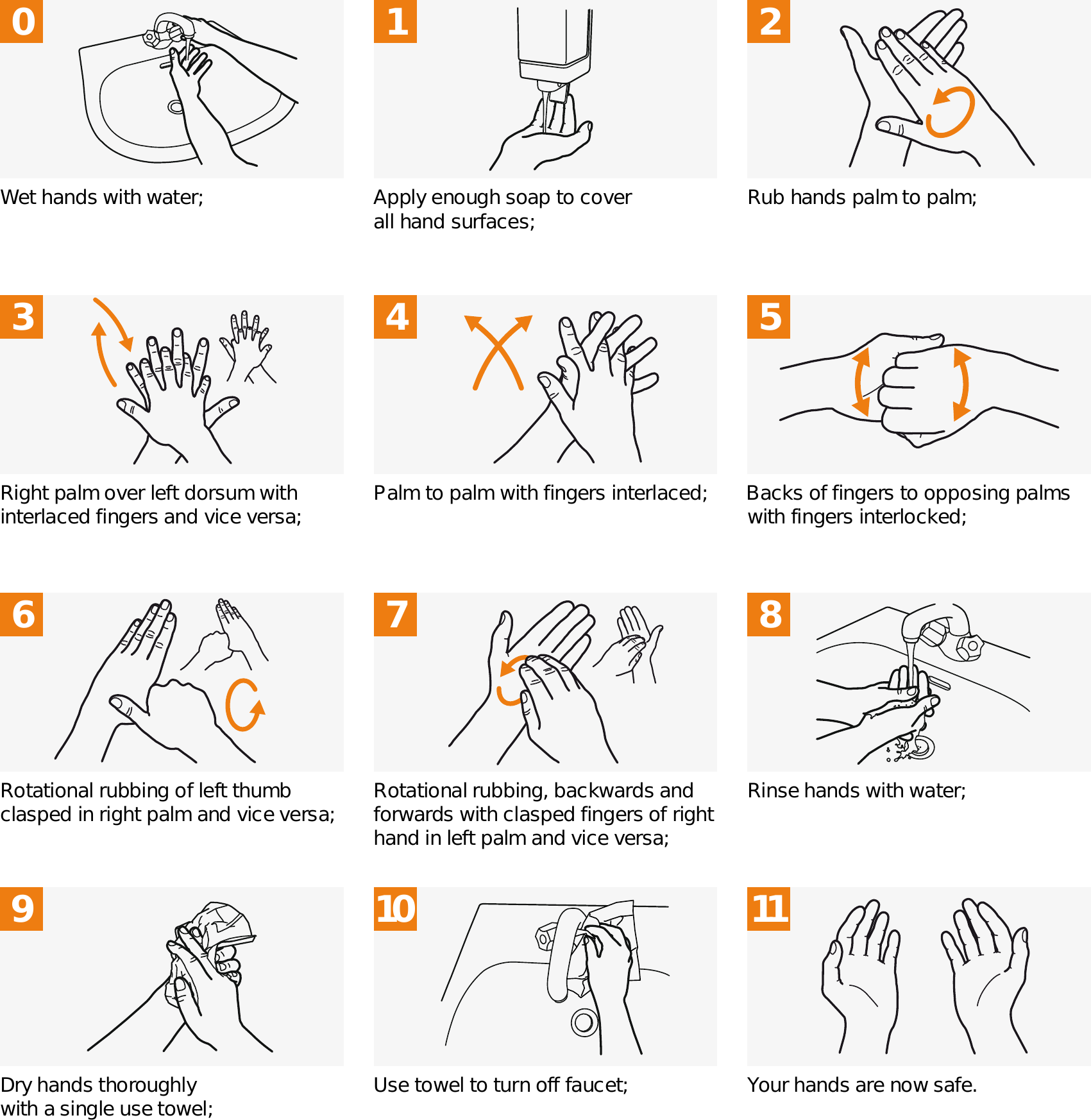}
    \caption{Hand hygiene technique with soap and water according to the WHO~\cite{world2009guidelines}. The work in this paper aims to automatically recognize the movements 2--7 and 10.}
    \flab{who}
  \end{center}
\end{figure}

Washing hands is one of the most important ways to prevent infectious diseases, including COVID-19.
The World Health Organization (WHO) has published hand washing guidelines~\cite{world2009guidelines} (Fig.~\fref{who}).
Unfortunately, even medical professionals often fail to follow these guidelines~\cite{widmer2004alcohol, widmer2007introducing, sutter2010effect, szilagyi2013large}, which highly increases the chances of spreading infections.
In the pre-COVID years, it was estimated that there were 4.1 million preventable medical-care related infections annually in Europe, which cost 37\,000 lives annually~\cite{eu-data}.


As a result, it is necessary to introduce automated quality control of hand washing in clinical environments with the goal to provide feedback on the performance of medical professionals. One of the most efficient ways of doing it is to deploy computer vision systems. 
Hand gesture recognition has already been applied to monitor hand hygiene and hand-washing compliance of medical staff\cite{hoey2007assisting, llorca2011vision,ward2014automated,galluzzi2015automatic}.
However, these systems typically evaluate the washing with a simple binary metric (done/not done) or use a single timer to detect its total duration. Neither is sufficient to ensure proper cleaning of all palmar surfaces~\cite{ward2014automated}.
To the best of our knowledge, it is not currently possible to automatically check whether the hand washing guidelines defined by the WHO are being followed, despite the ubiquity of these (or similar) guidelines. Such a situation is unsatisfactory, as while compliance with the guidelines is necessary to ensure safety of both patients and medical staff, medical professionals may forget, avoid performing, or cut short these movements in real-world clinical settings due to their workload and lack of time.

The solution to this problem is automated quality control with real-time feedback, which would be useful both for teaching the correct movements, and reminding about them in clinical practice.
In this paper, we present such an automated system that accounts not only for the total time, but also for the duration of each of the different washing movements as defined by the WHO.

To recognize gestures of hand-washing, the system uses deep neural network (DNN) models.
DNNs are multilayer assemblies of nonlinear units known as artificial neurons; in the last decade, DNNs have achieved state-of-the-art results on a variety of tasks, especially those dealing with perceptual data such as images. We use DNN models MobileNetV2 \cite{howard2017mobilenets} and Xception \cite{chollet2017xception} pretrained on Imagenet dataset \cite{deng2009imagenet}, which are available in Keras deep neural network library \cite{chollet2015keras}. 
As Keras uses TensorFlow tensor library \cite{abadi2016tensorflow} as its deep learning backend engine, it will be possible in the future to export these models trained on our dataset for the deployment on mobile and edge devices.

The neural networks are trained on a part of a large hand washing dataset collected in this project and presented in this paper.
The full dataset\footnote{\url{https://makonis.edi.lv/s/oEzC6n5rQi6Sd5Y}} includes 1854 annotated video files, 1094 of which were annotated by two persons for additional reliability.
Each video file includes a single hand washing episode.
The dataset was collected in a local hospital using data collection systems placed above water sinks in nine different locations.

The paper makes the following contributions:
\begin{itemize}
\item We present a large labeled real-world dataset of hand washing episodes;
\item We provide preliminary results of training and evaluating neural network models based on the state-of-the-art MobileNetV2 and Xception models;
\item We describe an application for hand washing quality control and feedback that is based on our neural network model.
\end{itemize}

The paper is structured as follows:
Section~\sref{relwork} overviews the related work in the field;
Section~\sref{collection} describes our real-world dataset of hand washing movements and its collection;
Section~\sref{deeplearning} describes and evaluates the neural network for hand washing movement recognition;
Section~\sref{application} presents the mobile device application for hand washing quality control and feedback.


%% file: 02-relwork.tex
\section{Related work}
\slab{relwork}

Ward \etal~\cite{ward2014automated} provide a systematic assessment of automated hand hygiene monitoring systems based on 42 scientific articles. However, fewer than 20\,\% of the surveyed articles include calculations for efficiency or accuracy of the presented systems, leaving these topics as an area where future studies are required~\cite{ward2014automated}. Issues with cost, patient privacy, and lack of validation were also identified. Some of the surveyed systems are classified as ``fully autonomous''; however, these systems include a wearable and mobile component. Our design aims for a fully autonomous system that does not require any extra effort from the medical staff such as wearing a personal wristband device.

Srigley \etal~\cite{srigley2015hand} provide a systematic efficacy review of hand hygiene monitoring.
Most of the studies they survey only monitor general compliance to hand hygiene guidelines; only one study also monitors the rate of hand hygiene events. In terms of feedback, only two systems provided individualized feedback and real-time reminders. The evidence for clinical adoption of hand hygiene monitoring is deemed insufficient; however, a suggestion to focus on video-based monitoring is made.

A number of inertial sensor-based approaches exist. Wang \etal~\cite{wang2020accurate} uses armband sensor data and achieve high recognition accuracy of the different movements in the WHO guidelines (Fig.~\fref{who}), especially for the user-dependent model (96\,\%, \emph{vs.} 82\,\% for user-independent model). Galluzzi \etal~\cite{galluzzi2015hand,galluzzi2015automatic} report 93\,\% accuracy using wrist-worn accelerometers when recognizing the movements defined by the WHO. Commodity devices are used, but they are worn on both wrists.

In terms of vision based systems, one of the first approaches is described in Hoey \etal~\cite{hoey2007assisting}. Their target application is an assistant for hand washing for dementia patients. They implement a particle filter-based classification approach and provide real-time feedback to the user. However, they do not aim to recognize different types of washing movements.

Llorca \etal~\cite{llorca2011vision} present another vision-based system with the explicit goal to provide an automatic hand washing quality assessment and recognize six different washing ``poses''. As research that predates the deep learning era, it uses a classical machine learning approach with a complex pipeline involving skin color detection, hand segmentation, and a particle filtering model for hand tracking. The motions are recognized by a Support Vector Machine (SVM) classifier. Even though the reported accuracy is high (70.1\,\% to 97.8\,\% depending on the motion on a dataset with 4 test subjects), the generalizability of the approach is under question for \eg more test subjects, especially ones with darker skin color, and for videos taken in real-world conditions.

In a more recent paper, Yeung \etal~\cite{yeung2016vision} present a deep-learning based vision-based system for hand hygiene monitoring. The system uses depth sensor modality instead of the full video data to preserve privacy. The data is classified using a convolutional neural network (CNN). However, the authors aim to recognize adherence to hand hygiene in general, instead of zooming in to the level of specific hand movements as done in the present work.

A study by Li \etal~\cite{li2019hand} investigates CNN for gesture recognition in general and achieves high accuracy, showing that this approach is suitable for the task.

Yamamoto \etal~\cite{yamamoto2020classification} use vision-based systems and a CNN for a different problem -- namely, the estimation of how well the hands are washed. They compare the quality score of the automated system with a ground-truth data obtained using a substance fluorescent under ultraviolet light. Their results show that the CNN is able to classify the washing quality with high accuracy.



%% file: 03-datacollection.tex
\section{Data Collection}
\slab{collection}


\begin{figure*}
  \begin{center}
    \centering
    \begin{subfigure}{0.61\textwidth}
      \includegraphics[width=\linewidth]{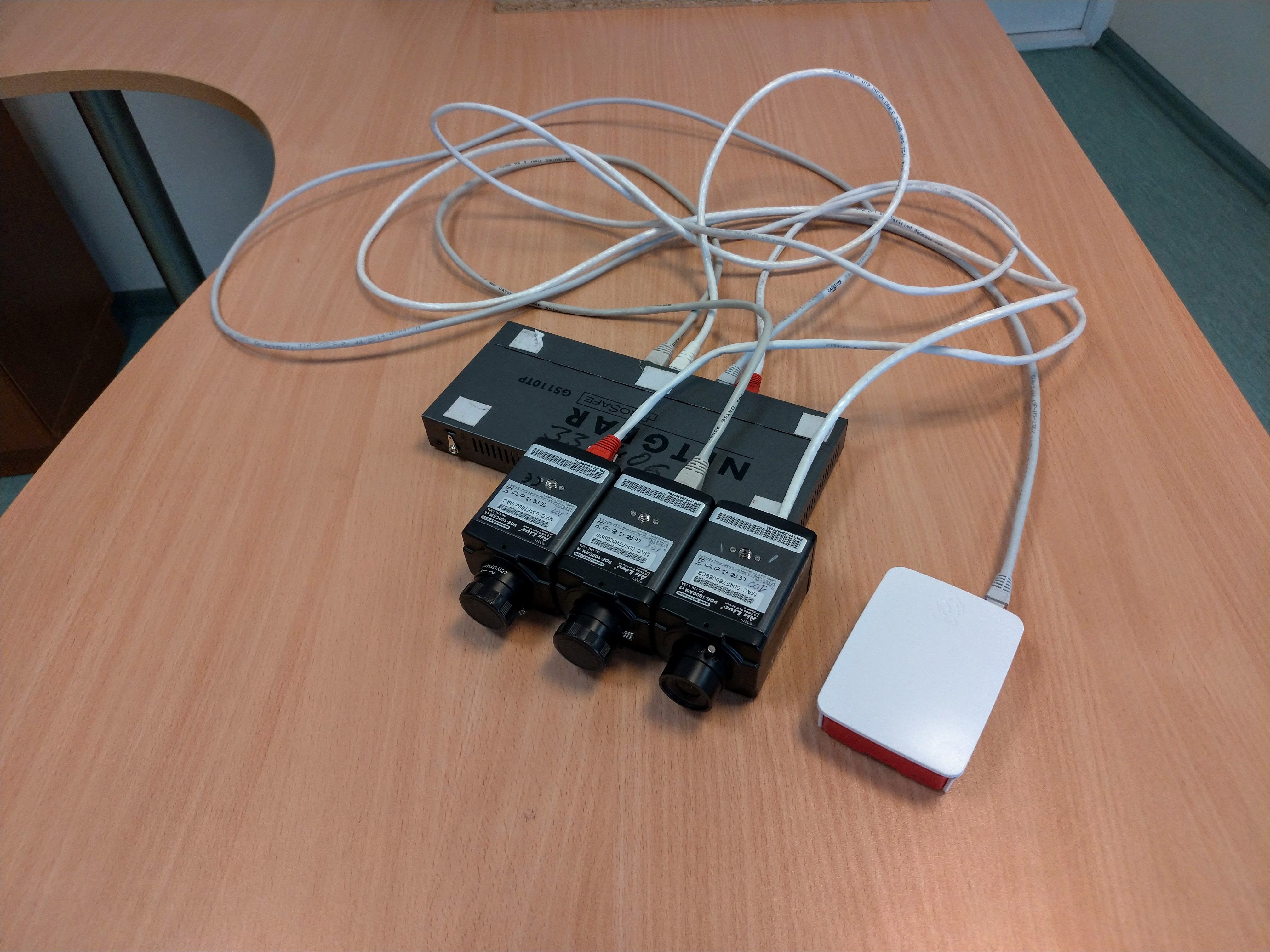}
      \caption{On-table view.}
      \flab{system-table}
    \end{subfigure}
    \hspace{0pt}
    \begin{subfigure}{0.374\textwidth}
      \includegraphics[width=\linewidth]{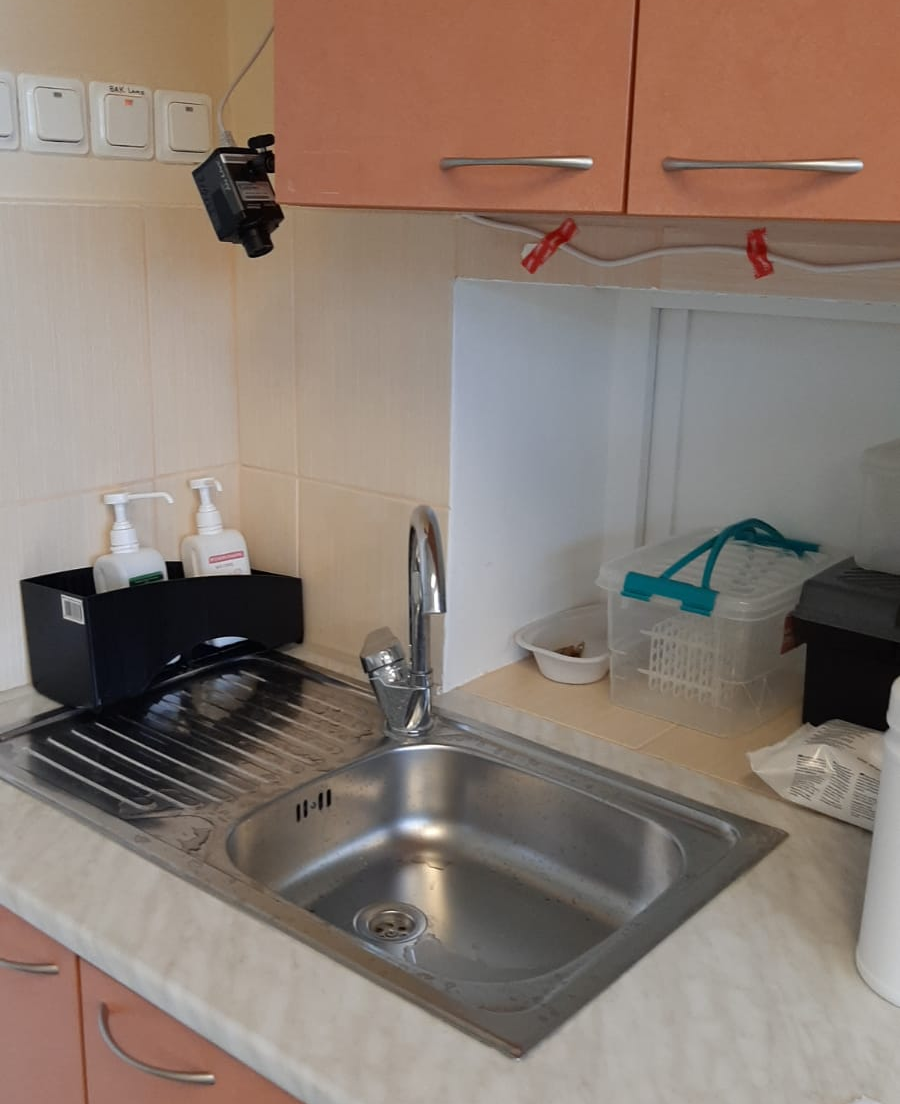}
      \caption{Deployed in a hospital.}
      \flab{system-hospital}
    \end{subfigure}
    \caption{Data collection system.}
    \flab{collection-system}
  \end{center}
\end{figure*}


\subsection{Data Collection System}

In order to train the neural network, the necessary data (video files with hand washing episodes) were collected and annotated according to hand washing guidelines recommended by the WHO.
For the collection of the data, we designed a custom IoT system (Fig.~\fref{system-table}).
Video files were captured using AirLive IP cameras POE 100CAM and Axis M3046V IP cameras, controlled by Raspberry Pi 4 single-board computers.
Recording parameters were set to following: video resolution 640x480 pixels, frame-rate 30 frames per second.
Recordings with the hand washing episodes were captured and stored on a micro SD card only when motion with a duration greater than 10\,sec was detected in front of the camera.
The complete data acquisition system consists of one or multiple IP cameras connected to a Netgear 5-Port PoE Gigabit Ethernet switch GS305P, and a Raspberry Pi 4 device with a micro SD card.
The micro SD card includes Raspberry operating system, our custom data acquisition program, and free space for storing acquired video files.
These data acquisition systems were installed above the sinks, at nine sites at Pauls Stradins Clinical University Hospital (Fig.~\fref{system-hospital}).
The data from the deployed systems was collected monthly, and the total deployment duration was 3 months.

\subsection{Data Annotation}

Subsequently, the collected video files were stored and annotated.
The purpose of the annotation was to match each specific hand washing movement to a corresponding code defined in our annotation guidelines.
To carry out the process, we developed a custom annotation program using the Python programming language and the OpenCV computer vision library (Fig.~\fref{annotation}).
The program allows to assign each video frame a corresponding hand washing movement.

\begin{figure}[b]
  \begin{center}
    \includegraphics[width=0.5\linewidth]{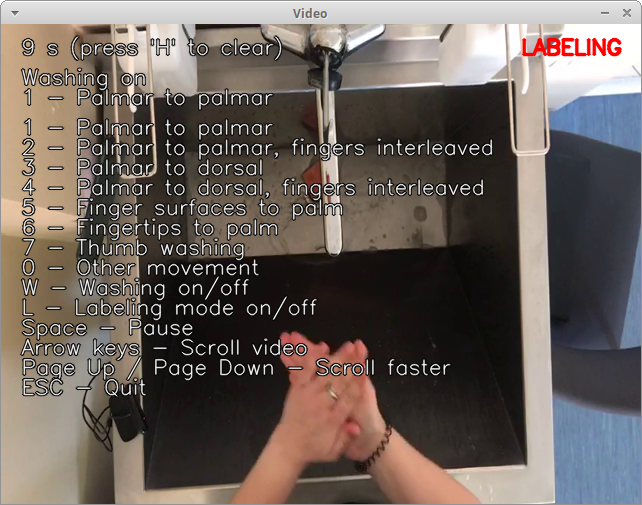}
    \caption{Annotation app.}
    \flab{annotation}
  \end{center}
\end{figure}

The annotation guidelines were developed in collaboration with local infectious disease specialists.
In total, we defined seven different hand washing movements as recommended by the WHO.
These movements are as follows: palm to palm, palm over dorsum with fingers interlaced, palm to palm with fingers interlaced, back of fingers to opposing palm, rotational rubbing of the thumb, fingertips to palm, turning off the faucet with a paper towel.
It was also important to annotate whether the person who is washing hands is wearing a ring, watch, or has lacquered nails, as all these aspects interfere with the best practices of hand washing and can be considered inappropriate for medical professionals in their work environment.
The videos were annotated by the infectious disease specialists involved in the project, medical professionals, as well as volunteers including Riga Stradins University students.

\subsection{Dataset Format}

The dataset (see Table~\tref{dataset}) consists of a number of video files, each of which corresponds to a single hand washing episode.
For each video file, there is a matching \code{.json} file containing the annotations in JSON format, and a \code{.csv} file containing the statistics of the video in comma-separated value format.

In order to improve the reliability of the dataset, some of the video files were annotated multiple by two different annotators. This approach will provide us an opportunity to compare and evaluate whether the performance of a neural network is more accurate when trained on a larger set of data, or on a smaller, but more accurately annotated one.

\begin{table} 
  \caption{Dataset summary statistics}
    \begin{center}
      \tlab{dataset}
      \normalsize
      \begin{tabular}{lc}
        \toprule
        \textbf{Parameter} & \textbf{Value}\\
        \midrule
        Total video files & 32471 \\
        Total annotations & 3387 \\
        Total annotated files & 2293 \\
        Files annotated 1 time & 1199 \\
        Files annotated 2 times & 1094 \\
        \bottomrule
      \end{tabular}
      \vspace{0pt}
    \end{center}
\end{table}


%% file: 04-recognition.tex
\section{Deep Learning for Washing Movement Recognition}
\slab{deeplearning}

Preliminary experiments with hand washing movement recognition were conducted with two convolutional neural networks, MobileNetV2 \etal~\cite{howard2017mobilenets}, and Xception \etal~\cite{chollet2017xception}. MobileNetV2 is a compact and fast CNN, whereas Xception is a larger CNN offering higher precision at the cost of slower training and inference speed. Both these networks are available in Keras \etal~\cite{chollet2015keras}, a high-level deep learning API for the deep learning tensor library TensorFlow \etal~\cite{abadi2016tensorflow}. The advantage of Keras is that it makes easy to access, modify, and train DNN models, whereas TensorFlow makes working with models fast and allows to deploy trained models on mobile and edge computing devices, which is envisaged as an extension of the present work.

\begin{figure*}[h]
  \begin{center}
    \includegraphics[width=0.8\linewidth]{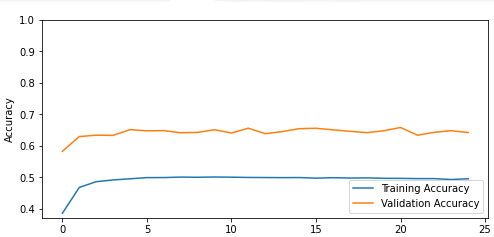}
    \caption{Training and validation accuracy of MobileNetV2 model.}
    \flab{MobileNetV2_accuracy}
  \end{center}
\end{figure*}

To obtain the first results promptly, we employed transfer learning approach, i.e. used MobileNetV2 and Xception models pretrained on the Imagenet dataset \etal~\cite{deng2009imagenet} rather than trained them from scratch. The dataset for training, validation, and testing our models was a part of the dataset described in Section~\sref{collection} and consisted of 378 video files. After splitting these video files into separate frames, 309315 frames were obtained, 70 per cent of which were used for training the models, 20 per cent for validation, and 10 per cent constituted the test set. The frames were resized to the maximum input sizes for the implementation of MobileNetV2 and Xception models in Keras, which are respectively 224x224 pixels and 299x299 pixels. Furthermore, a standard image augmentation procedure was applied to the images, namely, they were randomly flipped and rotated by 20 degrees. 

To obtain the preliminary results, MobileNetV2 was trained for 50 epoch with early stopping after 10 epochs if there was no performance improvement on the validation set for 10 epochs. No hyperparameter tuning was done; the model were trained using the categorical loss function and RMSprop optimization algorithm with the learning rate of 0.0001. As a result, performance of MobileNetV2 stopped to improve after the third epoch (Fig.~\fref{MobileNetV2_accuracy}), and the model achieved classification accuracy of 0.6403.

Xception model was trained for 10 epochs using the categorical loss function and Adam optimization algorithm with the default learning rate of 0.001. As a result, classification accuracy of 0.6683 was achieved.

%% file: 05-application.tex
\section{Application}
\slab{application}

The neural network (Section~\sref{deeplearning}) is a core component of the automated quality control and feedback application.
The application is intended to be used on smartphones or tablets mounted above a sink in clinical environment (Fig.~\fref{application}).
The functionality of the application includes continuously running video camera in active mode and monitoring the camera input for movement.
As soon as a movement is detected, classification using the neural network is started.
The application detects the duration of each movement and the total duration of the washing episode.
A washing episode is deemed to be successfully completed when the total duration exceeds a pre-configured threshold and all required washing movements have been detected.
The interface between the neural network and the rest of the application is implemented via periodic polling.
Each polling event reads two variables from the model: washing status (on/off) and the washing movement number (an integer corresponding to the movement number in the WHO guidelines).

\begin{figure*}[b]
  \begin{center}
    \includegraphics[width=\linewidth]{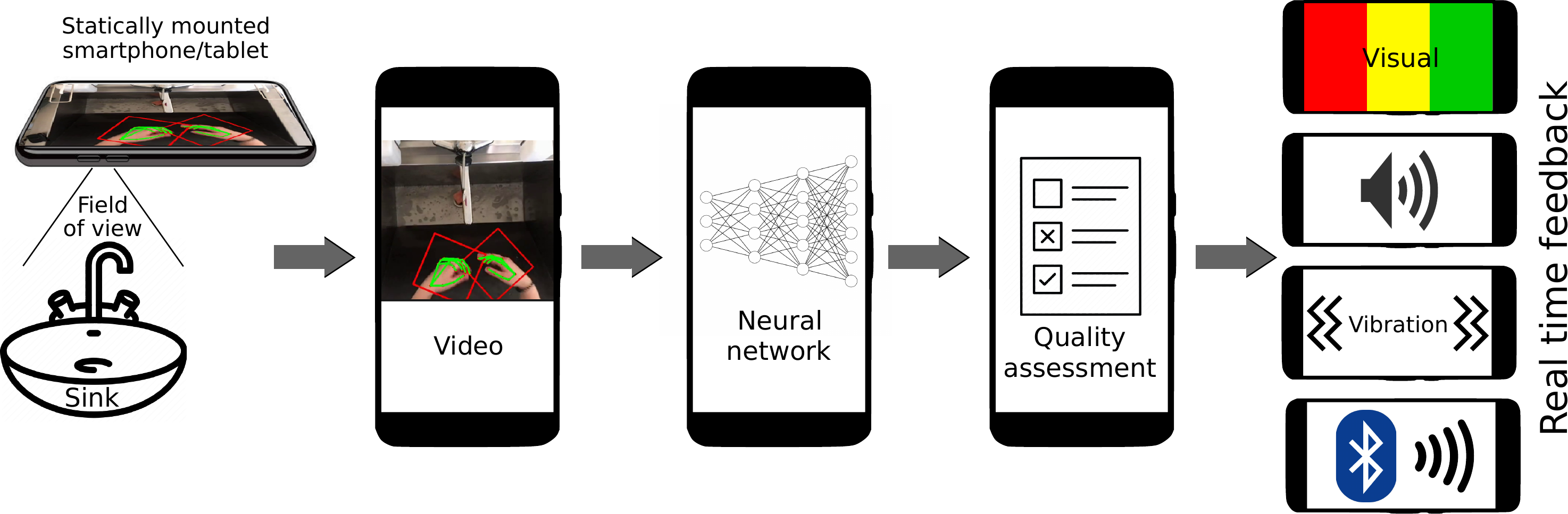}
    \caption{Monitoring and Feedback Application Concept.}
    \flab{application}
  \end{center}
\end{figure*}

\begin{figure*}
  \begin{center}
    \includegraphics[width=\linewidth]{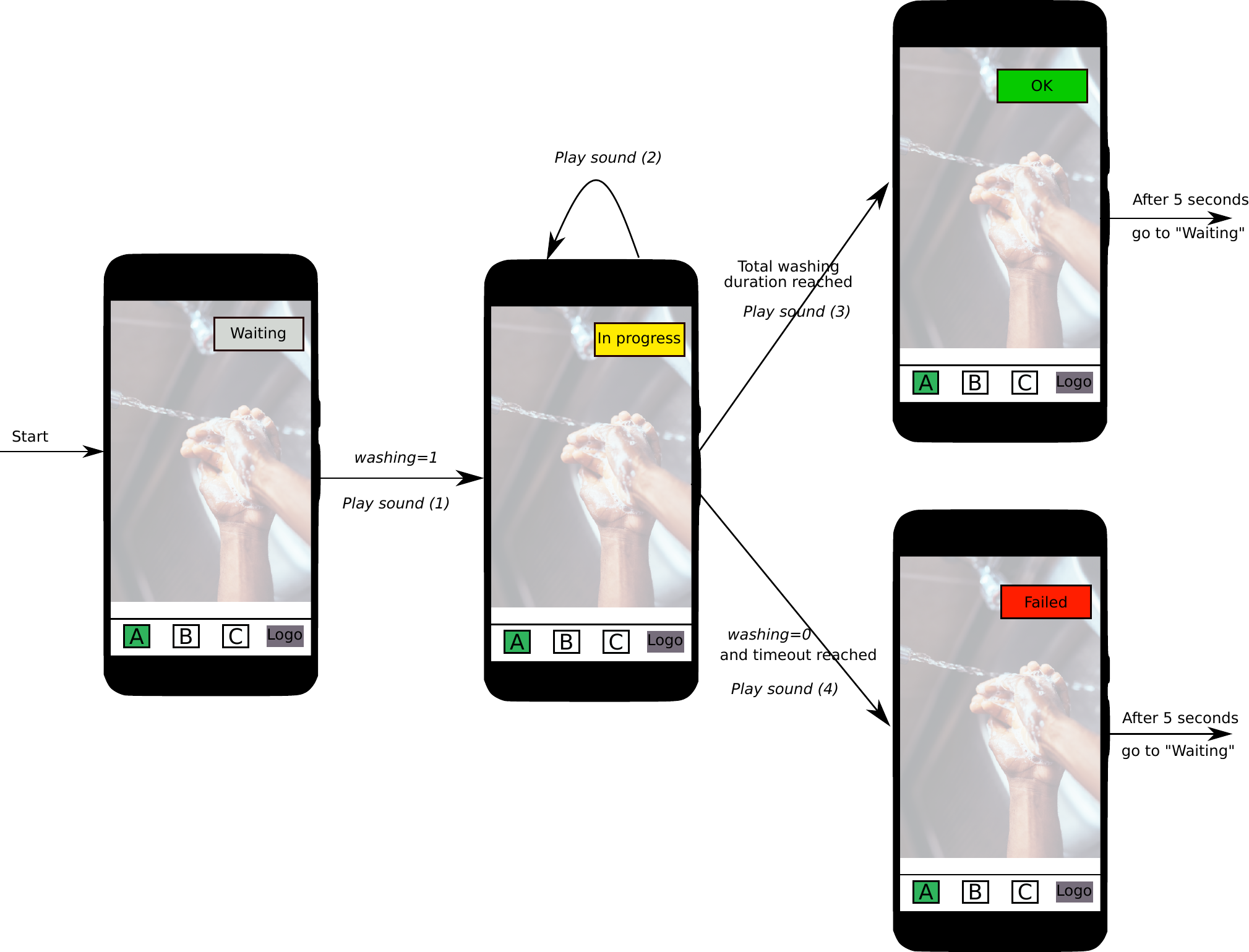}
    \caption{Application State Machine.}
    \flab{appstates}
  \end{center}
\end{figure*}

More formally, the logic of the application corresponds to a state machine with the following states (Fig.~\fref{appstates}):
\begin{itemize}
\item \emph{Waiting state.} The application watches for the start of the washing activity.
\item \emph{In-progress state.} The washing is on-going. The application keeps tracks of the duration of the washing, as well as about which movements have been present so far.
\item \emph{Ok state.} The washing is on-going, but total duration has been reached and all required movements have been present.
\item \emph{Failed state.} The washing has ended before reaching the total duration or before all movements were present.
\end{itemize}

It’s worth stressing that such a system, as envisioned by us, will be adaptable to changes in the WHO guidelines.
If the duration of the existing washing movements or the total duration of the washing is changed, the system's core neural network does not have to be changed; changing the configurable system parameters is enough.
The neural network only must be retrained if new movements are added, which is not likely given the longevity of the WHO guidelines.
